\documentclass{moriond}

\def\be{\begin{equation}}
\def\ee{\end{equation}}
\def\bea{\begin{eqnarray}}
\def\eea{\end{eqnarray}}

\newcommand{\Photo}{\includegraphics[height=35mm]{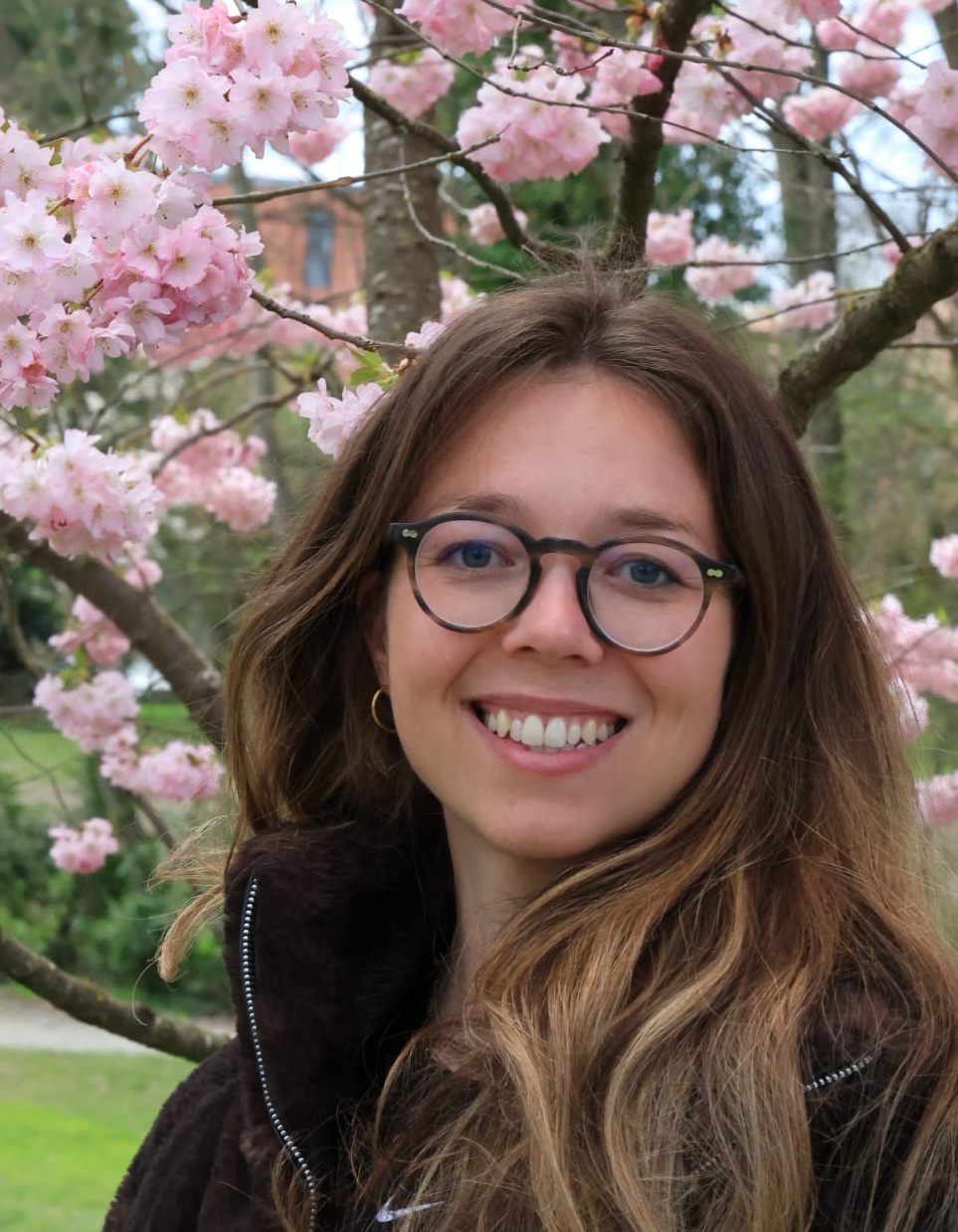}}

\begin{document}
\vspace*{4cm}
\title{Electroweak physics and long-lived particles at LHCb}

\author{ Felicia Volle on behalf of the LHCb Collaboration }

\address{School of Physics and Astronomy, University of Birmingham, Edgbaston, 
Birmingham B15 2TT, United Kingdom}

\maketitle\abstracts{
Extensions of the Standard Model (SM) of Particle Physics can be probed either through precision measurements of SM observables or via direct searches for processes beyond the SM (BSM). This proceeding focuses on precision measurements in the electroweak sector, in particular the properties of the $Z$ boson, $W$ boson and top quark. Measurements of the $W$ and $t$ production cross-sections, as well as charge asymmetries, with an integrated luminosity of 5.1 and 5.4~fb$^{-1}$ of $pp$ collisions collected by the LHCb experiment, are presented for the first time. In consequence of the forward coverage of the LHCb detector, these results provide complementary probes on parton distribution functions compared to measurements performed at central rapidity. Well-motivated BSM candidates include mediators between the visible and dark sectors. In this context, recent results from searches for axion-like particles and heavy neutral leptons are also discussed.
}

\section{Probing the Standard Model}

The completeness of the Standard Model (SM) of particle physics can be tested through both direct and indirect approaches. Indirect searches rely on precision measurements of SM observables, where contributions from beyond the Standard Model (BSM) physics may contribute via higher-order corrections. Direct searches, in contrast, aim to observe BSM particles produced in the detector. In this proceeding, recent results from both approaches, published by the LHCb Collaboration, are presented.

\section{Electroweak measurements at LHCb}

Electroweak precision observables provide an excellent testing ground for BSM hypotheses through indirect probes. Measurements of production cross sections of, for example, the electroweak bosons ($W$ and $Z$) and the top quark also place important constraints on the parton distribution functions (PDFs) of the proton. In this context, the forward acceptance of the LHCb detector allows access to particularly low (and high) Bjorken-$x$ values~\cite{LHCb:2014set,Thorne:2008am}. This complementarity with measurements performed at central rapidity can have a significant impact when combined.

\subsection{\texorpdfstring{$Z$-boson mass measurements}{Z-boson mass measurements}}
\label{subsec:Zmass}

One of the key electroweak precision observables is the $Z$-boson mass. The LHCb Collaboration published its determination from $Z \to \mu^+\mu^-$ decays in the 2016 dataset of $pp$-collisions at $\sqrt{s} = 13$\,TeV~\cite{LHCb:2025nob}, which has an integrated luminosity of 1.7\,fb$^{-1}$. Since the mass is reconstructed from the final-state kinematics, a crucial ingredient of the measurement consists of an excellent detector calibration. Neglecting the muon mass, the $Z$-boson mass can then be expressed as: 
\begin{equation}
     m^2 \simeq 2 p_{\mu^{+}} p_{\mu^{-}}(1-\cos\theta). 
\end{equation}
While the opening angle between the two muons, $\theta$, is well determined, a precise momentum calibration is achieved using the $\Upsilon(1S)$ and $J/\psi$ resonances, as well as the pseudomass method at the $Z$ peak~\cite{Barter:2021npd,LHCb:2023yqm}. The background contributions are reduced to an order of $10^{-3}$. A $\chi^2$ fit to the dimuon mass distribution is then performed. The resulting value of the $Z$-boson mass is
\begin{equation}
    m_Z = 91185.7 \pm 8.3(\mathrm{stat}) \pm 3.9(\mathrm{sys}) \mathrm{ MeV}, 
\end{equation}
representing the first dedicated measurement of $m_Z$ at the LHC.

\subsection{\texorpdfstring{Model-independent $W$-boson mass measurement}{Model-independent W-boson mass measurement}}
\label{subsec:Wmass}

After the measurement of the $Z$-boson mass, a natural extension is the study of the $W$-boson mass, which provides complementary sensitivity to electroweak parameters. 
Motivated by recent tensions between the $W$-boson mass determination by CDF~\cite{CDF:2022hxs}, LEP~\cite{ALEPH:2013dgf} and other experiments, Ref.~\cite{LHC-TeVMWWorkingGroup:2023zkn} 
underlines that the combination is limited by the dependence on theoretical signal modelling, which is intrinsically built into the analyses.
A novel approach suggest the factorisation of experimental effects and signal modelling, such that the latter can be updated independently. Using a dataset corresponding to 100\,pb$^{-1}$ of $pp$ collisions at $\sqrt{s} = 5.02$\,TeV recorded by the LHCb experiment during a two-week long period in 2017, the $W$-boson cross section is extracted from $W^- \to \mu^- \bar{\nu}_\mu$ and $W^+ \to \mu^+ \nu_\mu$ decays as a function of the muon transverse momentum $p_{T}$~\cite{LHCb:2025msn}. Background separation is achieved via a template fit to the muon isolation variable, while no assumption has been made on the signal shape in muon $p_{T}$. 
These differential cross sections are corrected for detector effects, and are the basis to extract the $W$-boson mass, which is measured to be
\begin{equation}
    m_W = 80369 \pm 130 (\mathrm{exp}) \pm 33 (\mathrm{th}) \mathrm{ MeV}. 
\end{equation}
Note that this measurement does not supersede the LHCb result at $\sqrt{s} = 13$\,TeV~\cite{LHCb:2021bjt}, but is a proof-of-concept of this novel approach with an independent dataset.

\subsection{\texorpdfstring{$W$-boson production cross-section}{W-boson production cross-section}}

\begin{figure}[ht]
    \centering
	\includegraphics[width = 0.45\linewidth]{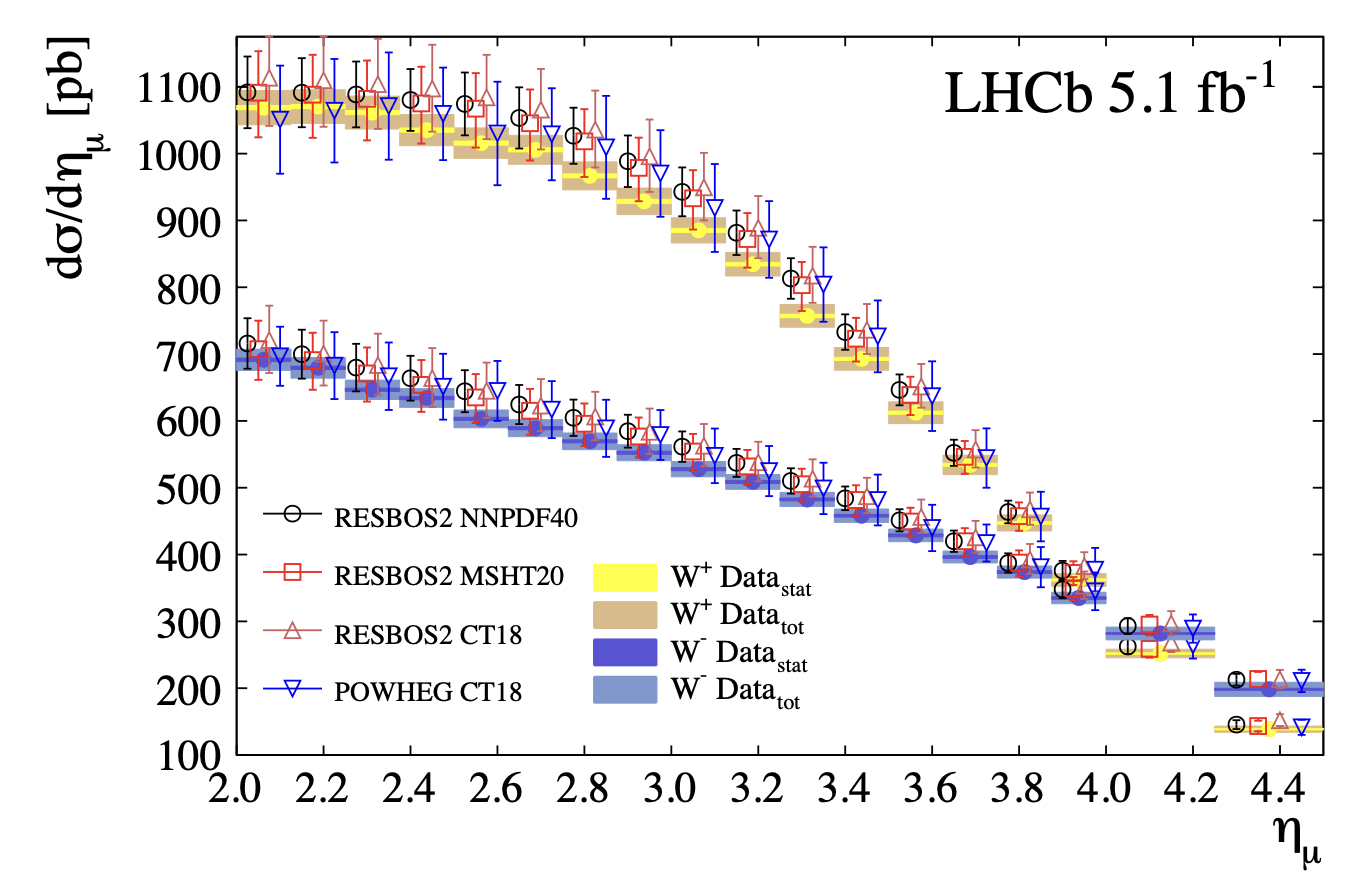}
	\includegraphics[width = 0.45\linewidth]{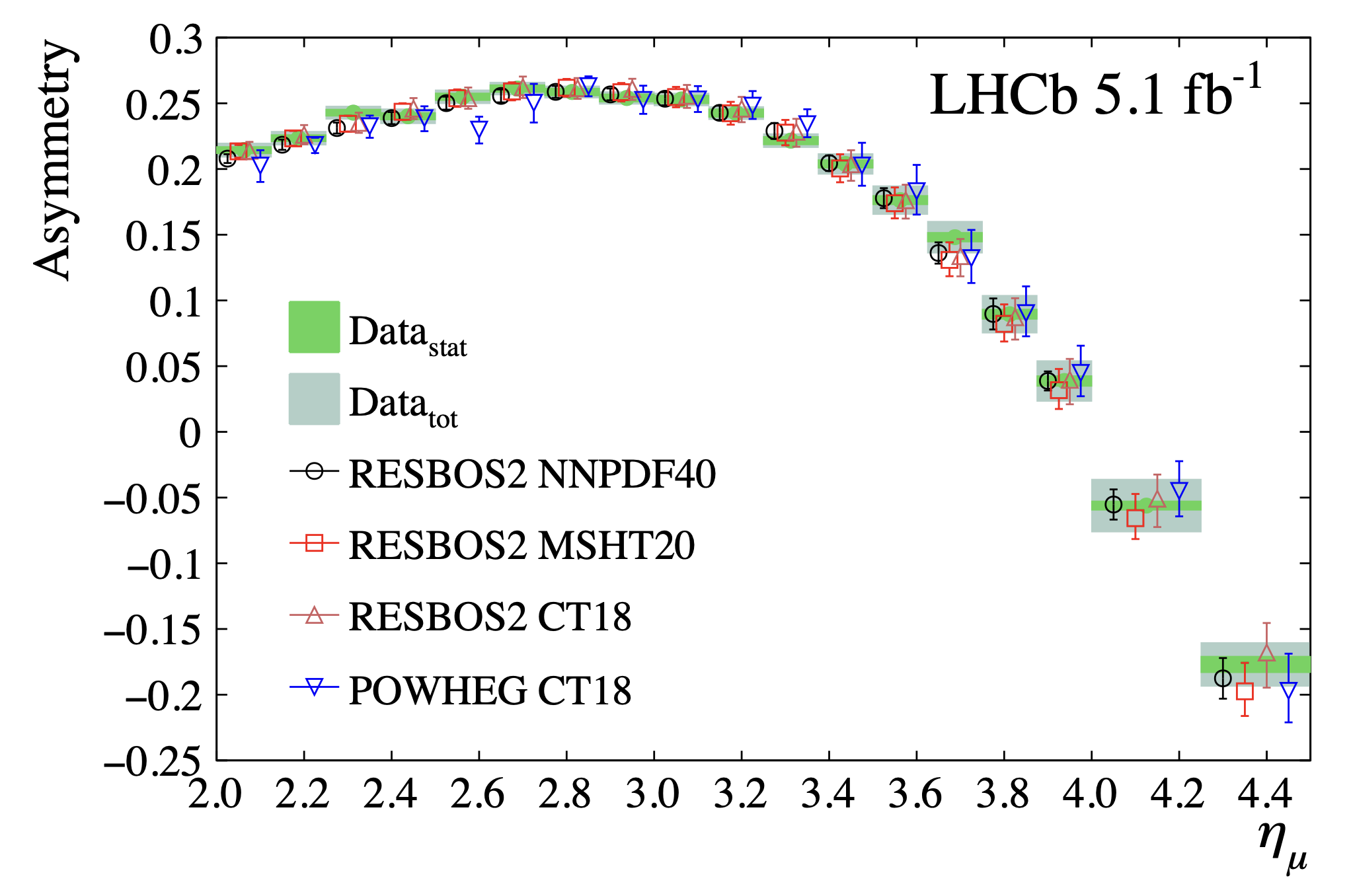}
    \label{fig:W-diff_13TeV}
    \caption{The differential $W$-boson production cross section\protect\cite{LHCb:2026dan} is drawn on the left as a function of the muon pseudo-rapidity, separated by the muon charge, and the muon charge asymmetry\protect\cite{LHCb:2026biq} shown on the right. Different generators and PDF sets are compared against the measurement, which bands correspond to the statistical (stat) and total (tot) uncertainty of 68\% confidence level.}
\end{figure}

The $W$ decay modes mentioned in Sec.~\ref{subsec:Wmass} do not only serve to determine the $W$-boson mass, but also to measure its production cross section. For this analysis, a dataset of $pp$ collisions at $\sqrt{s} = 13$\,TeV with an integrated luminosity of 5.1\,fb$^{-1}$ have been exploited~\cite{LHCb:2026dan}. As in Sec.~\ref{subsec:Zmass}, a momentum calibration is performed with the pseudomass method~\cite{Barter:2021npd,LHCb:2023yqm}, where $Z \to \mu^+\mu^-$ decays in data are utilised. The final template fits are performed in bins of the muon pseudo-rapidity $\eta_\mu$ separated by the muon charge. The muon $p_{T}$ serves to separate the signal process against the QCD background, electroweak and heavy flavour processes. Furthermore, corrections on the muon $p_{T}$ with respect to simulation and misidentification efficiencies are applied. The final differential cross section is shown on the left in Fig.~\ref{fig:W-diff_13TeV}. 

Thanks to the separation of the differential cross sections, the muon charge asymmetry~\cite{LHCb:2026biq} has been extracted as 
	\begin{equation}
		\mathcal{A}(\eta_\mu) =  \frac{d\sigma_{W^+}/d\eta_{\mu^+} - d\sigma_{W^-}/d\eta_{\mu^-}}{d\sigma_{W^+}/d\eta_{\mu^+} + d\sigma_{W^-}/d\eta_{\mu^-}}.
        \label{eq:MuonChargeAsym}
	\end{equation}
The resulting asymmetry as a function of $\eta_\mu$ is shown on the right in Fig.~\ref{fig:W-diff_13TeV}, which represents the most precise muon asymmetry determination in the forward region to date.

\subsection{Top quark production cross-section measurement}

\begin{figure}[h]
        \centering
		\includegraphics[width = 0.32\linewidth]{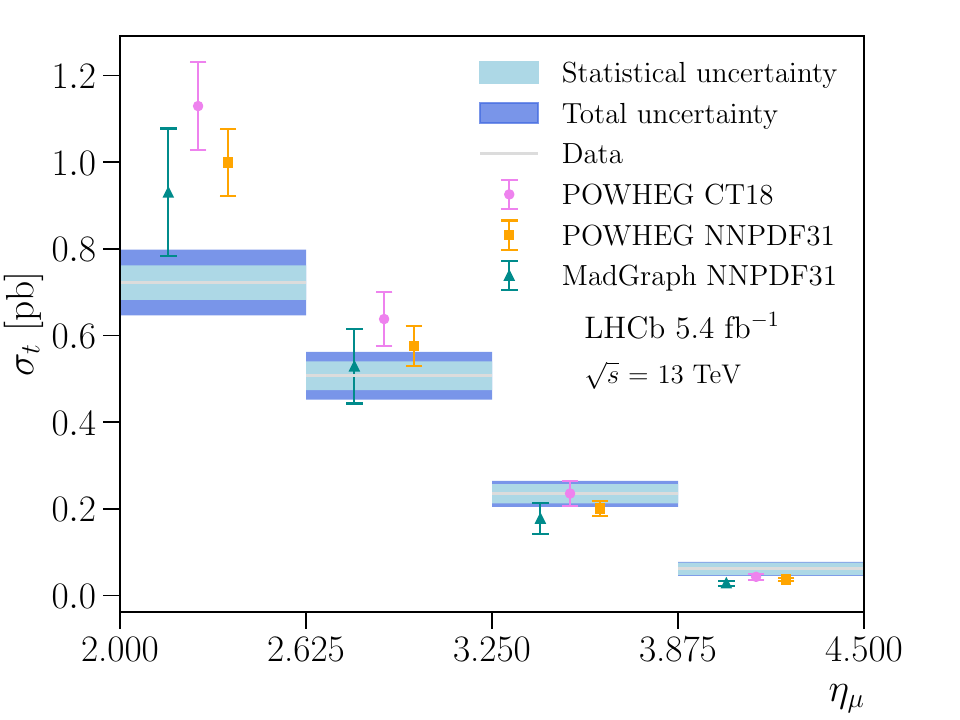}
		\includegraphics[width = 0.32\linewidth]{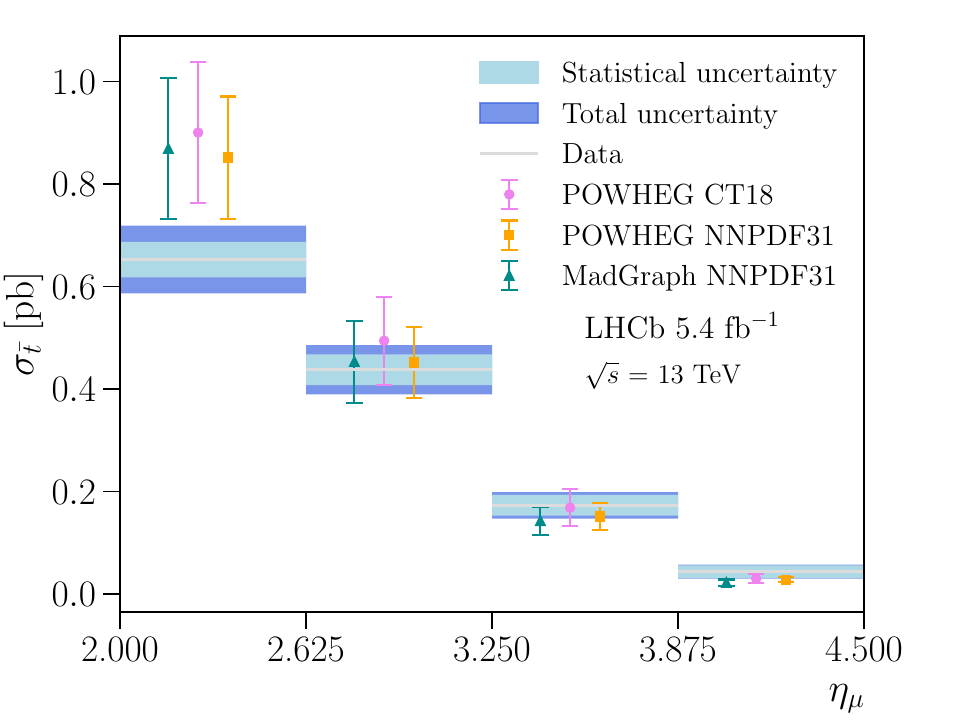}
		\includegraphics[width = 0.32\linewidth]{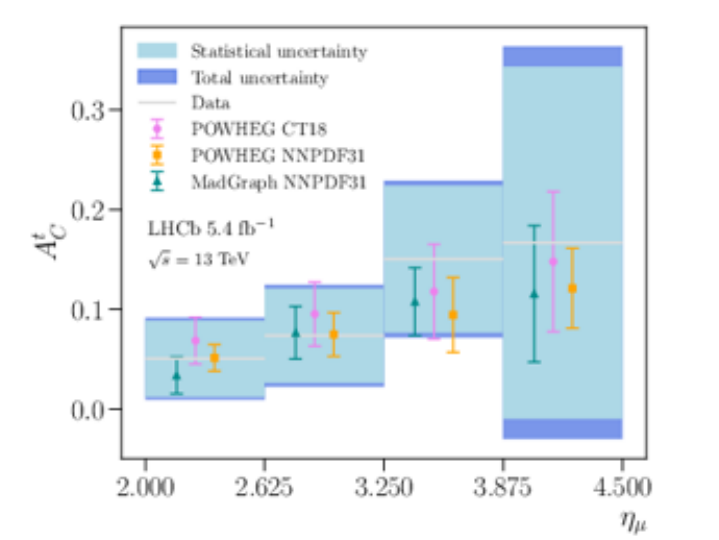}
        \label{fig:t_diff_cross}
        \caption{The differential production cross sections of the $t$ (left) and $\bar{t}$ quark (right) are drawn as a function of $\eta_\mu$~\protect\cite{LHCb:2025kfp}. Furthermore, the resulting top charge asymmetry~\protect\cite{LHCb:2025kfp} is shown. All the measurement are presented with the corresponding uncertainty bands, and compared to different generators and PDF sets.}
	\end{figure}

In a manner similar to the $W$-boson production cross section, the top-quark production cross section has been measured for the first time in the forward region by the LHCb Collaboration~\cite{LHCb:2025kfp}, which uses the $2016-18$ dataset of an integrated luminosity of 5.4\,fb$^{-1}$ collected in $pp$ collisions at $\sqrt{s} =13$\,TeV. The analysis exploits $t \to W^+(\to \mu^+\nu_\mu)b$ decay and its $CP$-conjugated counterpart. The $b$ quark has been reconstructed as a jet and clustered using the the anti-$k_T$ algorithm~\cite{Cacciari:2008gp} with a radius parameter of $R=0.5$. The jet flavour is determined by a multiclass neutral network, trained on a large set of input features related to the jet constituents and jet substructure~\cite{LHCb:2026ezb}. This new classifier improves the $b$- and $c$-tagging efficiencies by $11-53$\%, depending on the jet $p_T$, compared to the algorithm based on secondary vertex information~\cite{Freund:1997xna,Breiman:2017lcz,LHCb:2015tna}. Employing this tagging algorithm will be promising also for many other analyses, including Higgs-boson searches. A template fit is performed in bins of $\eta_\mu$ to separate backgrounds as a function of the $b$-jet probability $P_b$. The resulting differential cross sections are shown in Fig.~\ref{fig:t_diff_cross}.  

The forward region provides enhanced sensitivity to the muon charge asymmetry~\cite{Gauld:2014pxa}. Making use of the differential top-quark production cross sections, the muon charge asymmetry is extracted according to Eq.~\ref{eq:MuonChargeAsym}. The resulting distribution can be seen in Fig.~\ref{fig:t_diff_cross}. 

\section{Mediators between the visible and dark sector}

The second part of this proceeding is dedicated to direct searches for BSM physics. Well-motivated BSM candidates include mediators between the visible and the dark sector. While particles in the dark sector do not possess direct couplings to the SM, they can interact with it only via such mediators. The mediators considered in this proceeding are axion-like particles and heavy neutral leptons.

\subsection{Search for axion-like particles}

\begin{figure}[h]
    \centering
    \includegraphics[width=0.45\linewidth]{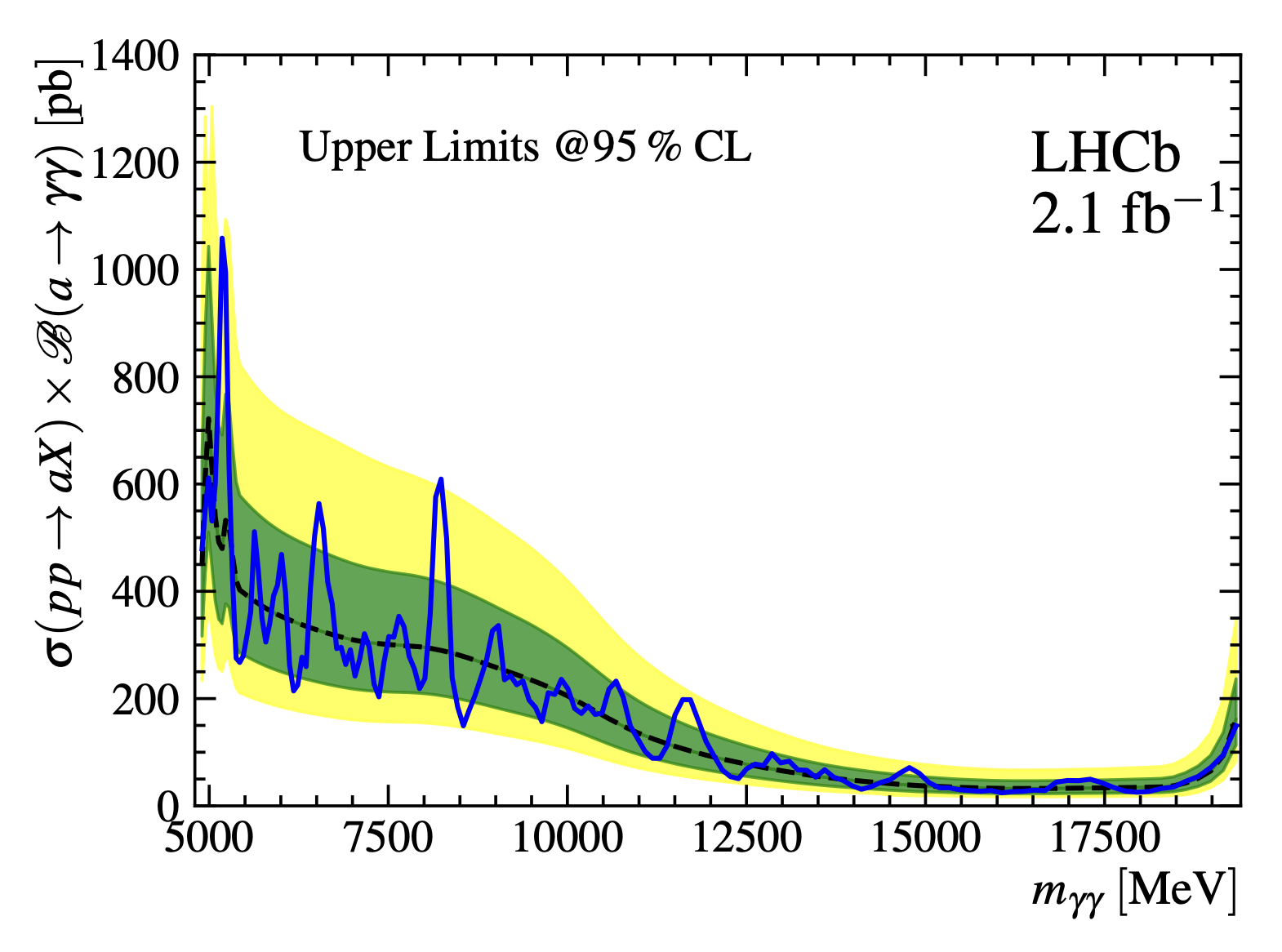}
    \includegraphics[width=0.45\linewidth]{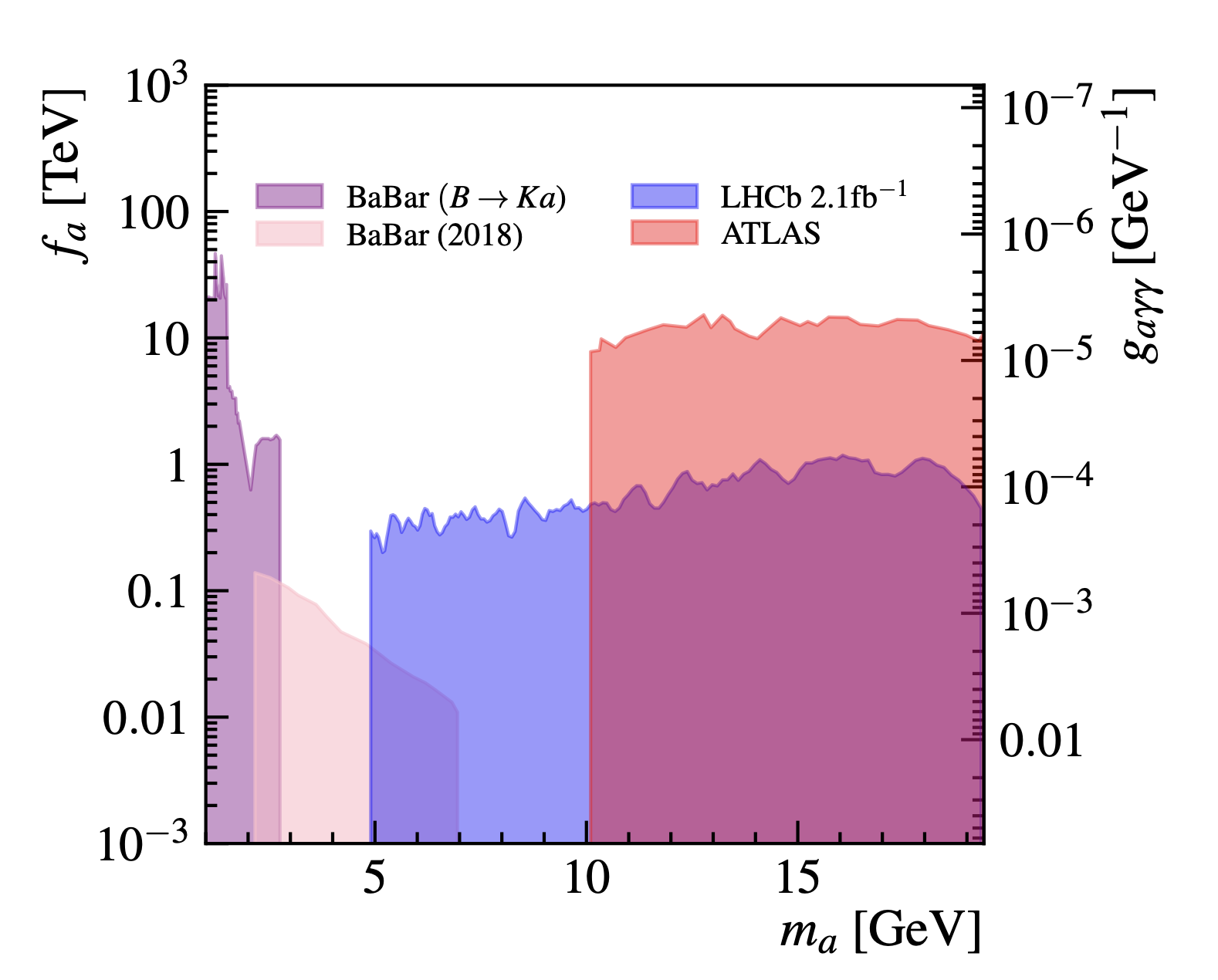}
    \caption{The upper limits at 95\% confidence level on the cross section times branching fraction are shown as a function of the ALP mass on the left. On the right, the limits are reinterpreted under the assumption $c_1 = c_2 = c_3 = 10$~\protect\cite{LHCb:2025gbn}.}
    \label{fig:upper_limits_alp2gg}
\end{figure}

Axion-like particles (ALPs) are generated as pseudo–Nambu–Goldstone bosons when a symmetry is spontaneously broken at an energy scale $f_a$. The coupling strengths of the ALP to the strong and electroweak $SU(3)_C \times SU(2)_L \times U(1)_Y$ interactions are expressed as $c_1$, $c_2$ and $c_3$, which are typically set to 10. 
A search for $\textnormal{ALP} \to \gamma\gamma$ decays has been performed in the 2018 dataset of $pp$ collisions collected by the LHCb experiment~\cite{LHCb:2025gbn}. The dominant coupling of the ALPs is to gluons, besides in the case of $c_3 \ll c_1, c_2$. Therefore, gluon-gluon fusion is the dominant production mode. The final-state of photons is motivated by its experimentally cleanness. The two photons are reconstructed under the assumption that they originate from the $pp$ collision. Multivariate methods against misidentification and signal isolation are employed. The combinatorial background and merged $\pi^0 \to \gamma\gamma$ decays remain and are modelled in the bump hunt, which is performed in steps of half of the diphoton mass resolution. No candidates have been found and the extracted upper limits on the cross section multiplied by the branching fraction at 95\% confidence level are shown on the left in Fig.~\ref{fig:upper_limits_alp2gg}. 
The upper limits have also been reinterpreted assuming $c_1=c_2=c_3=10$, which leads to the exclusion plot on the right in Fig.~\ref{fig:upper_limits_alp2gg}. The obtained results represent the best limits on $f_a$ for ALP masses in $[4.9, 10]$\,GeV and is the first search of a fully neutral final state with the LHCb experiment. 

\subsection{\texorpdfstring{Heavy Neutral Leptons in $B$-meson decays}{Heavy Neutral Leptons in B-meson decays}}

\begin{figure}[h]
    \centering
    \includegraphics[width=0.5\linewidth]{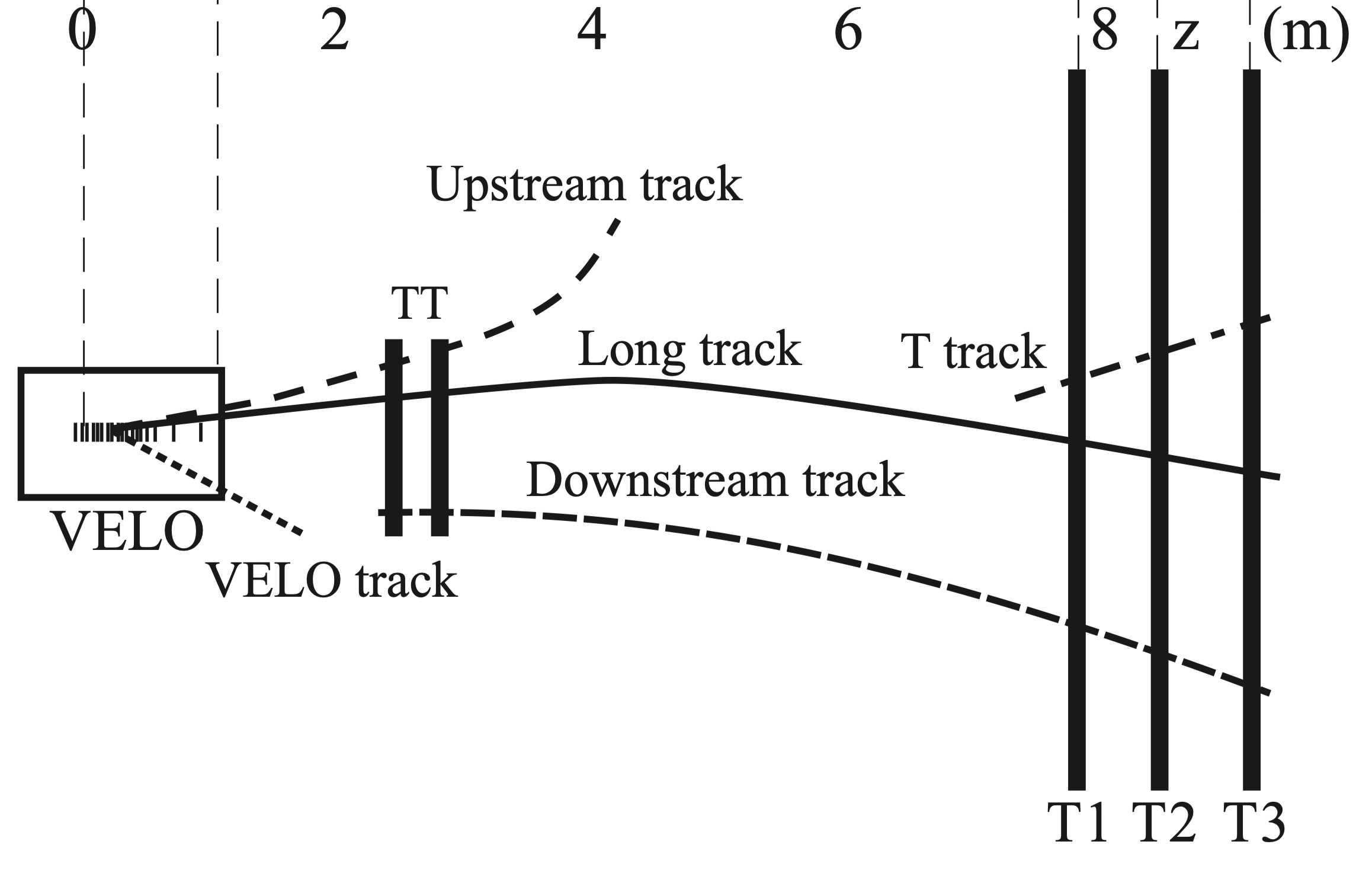}
    \caption{The different track types reconstructed at LHCb~\protect\cite{LHCb:2008vvz}.}
    \label{fig:track_types_lhcb}
\end{figure}

Heavy neutral leptons (HNLs) arise via mixing with the standard model neutrinos $\nu_\ell$, which can explain the suppression of the $\nu_\ell$ mass and a coupling of the HNL to $\nu_\ell$ via $|U_{\ell N}|^2$ arises. HNLs have been searched for in $B$-decays, namely the $B_{(c)}^ + \to \mu^+ N$ and $B \to \mu^+ NX$ decay modes in the $2016-18$ dataset of $pp$ collisions collected by the LHCb detector, corresponding to an integrated luminosity of 5.0\,fb$^{-1}$~\cite{LHCb:2025ymr}. The HNL decay mode considered in this search is $\mu^\pm \pi^\mp$. While Dirac-like HNLs decay only to muons with opposite charge with respect to the muon from the $B$ decay, Majorana-like HNLs can decay to both charge combinations. The search targets HNL masses in the range between 1.5 and 5.5\,GeV. The HNL can be reconstructed either inside or outside the vertex locator, corresponding to the \textit{long} and \textit{downstream} categories, respectively. The different track types are illustrated in Fig.~\ref{fig:track_types_lhcb}.

\begin{figure}[ht]
        \centering
    	\includegraphics[width=0.45\linewidth]{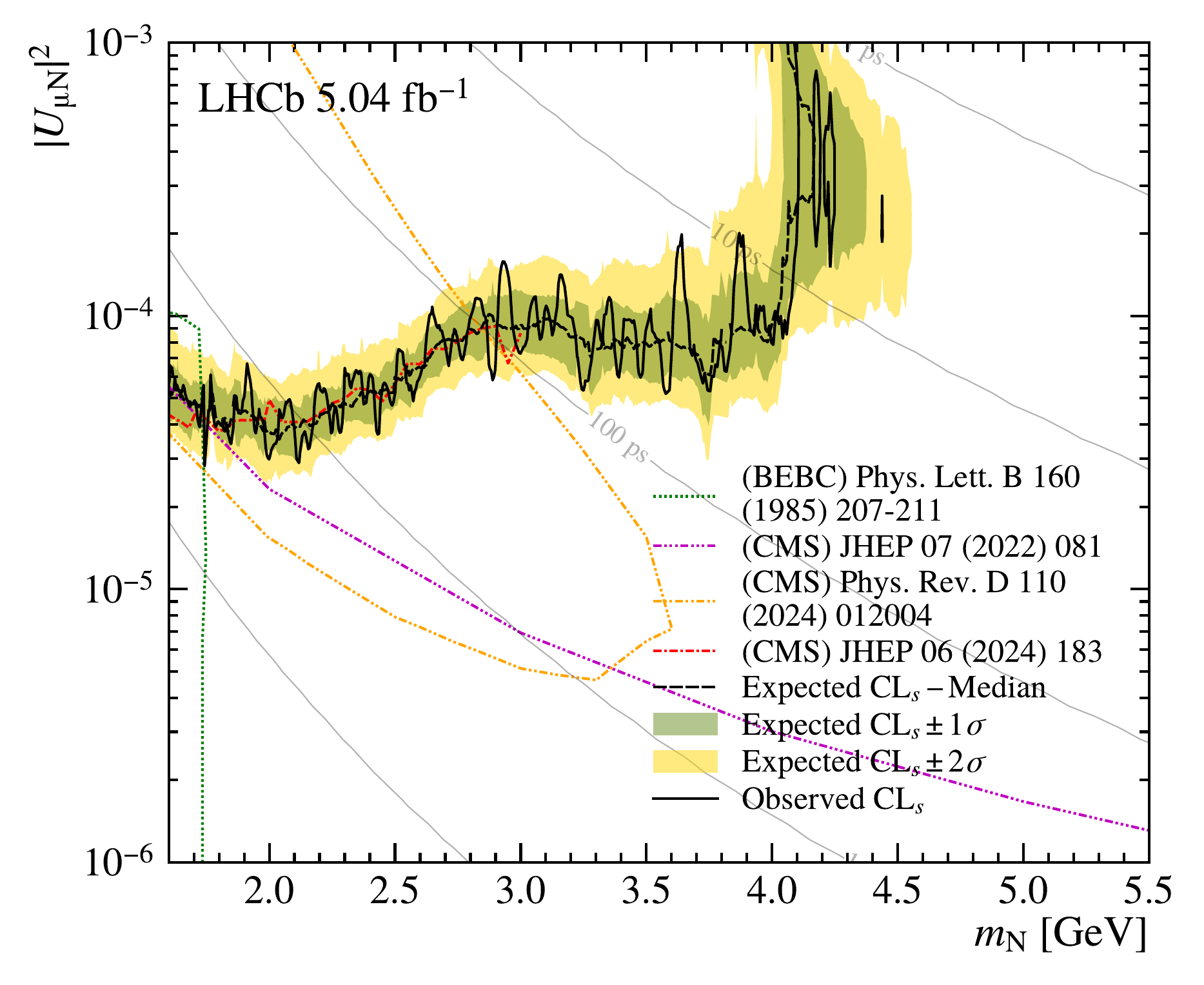}
        \includegraphics[width=0.45\linewidth]{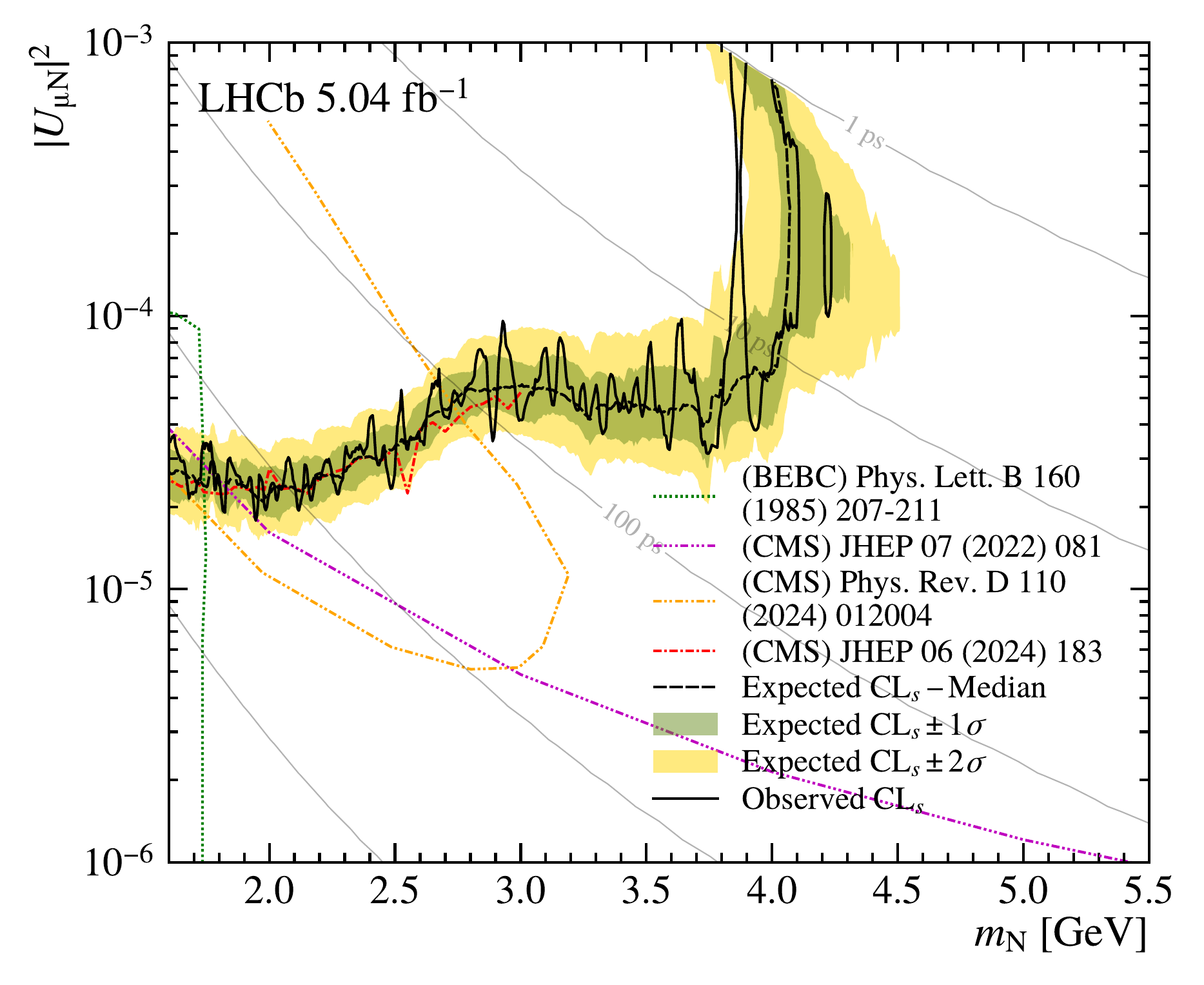}
        \caption{Upper limits on the coupling of the HNLs to muon neutrinos $|U_{\mu N}|^2$ at 95\% confidence level as a function of the HNL mass $m_N$ interpreted for Dirac-like (left) and Majorana-like (right) HNLs~\protect\cite{LHCb:2025ymr}.}
        \label{fig:HNL_limits}
	\end{figure}

A bump hunt is performed simultaneously to all the search categories. No candidate has been found and the resulting upper limits are presented in Fig.~\ref{fig:HNL_limits}. 
The obtained limits improved by an order of magnitude with respect to the Run~1 result published by the LHCb Collaboration~\cite{LHCb:2014osd}. 

Further improvements will be possible with the Run~3 dataset collected by the upgraded LHCb detector. The already acquired dataset, corresponding to 2.5 times the Run~1 and Run~2 integrated luminosity, will significantly increase the statistical power. Furthermore, additional long-lived signatures can be exploited, such as T-tracks, which allow the reconstruction of HNLs decaying after travelling approximately 2.5–8\,m in the $z$ direction. Further improvements can also be achieved by exploring additional HNL decay modes, such as $N \to \mu^- e^+ \nu_e$ and \mbox{$N \to \mu^- \mu^+ \nu_\mu$}. These developments indicate promising prospects for HNL searches at the LHCb experiment~\cite{LHCB-FIGURE-2025-001}.

\section{Conclusion}

In summary, this proceeding has presented recent results on electroweak precision measurements and BSM searches obtained by the LHCb Collaboration. New analysis tools have been exploited to achieve stringent results. One example is the improvement of $b$- and $c$-tagging efficiencies using a new tagger based on a deep neural network, which will impact a wide range of analyses. New analysis strategies have also been explored, such as a model-independent $m_W$ extraction and the use of fully neutral final states. Last but not least, new tracking strategies could enhance the sensitivity to long-lived signatures, thereby improving the discovery potential of BSM analyses.

\section*{References}
\bibliography{moriond}

@inproceedings{Thorne:2008am,
    author = "Thorne, R. S. and Martin, A. D. and Stirling, W. J. and Watt, G.",
    title = "{Parton Distributions and QCD at LHCb}",
    booktitle = "{16th International Workshop on Deep Inelastic Scattering and Related Subjects}",
    pages = "30",
    month = "8",
    year = "2008",
}

@article{LHCb:2014set,
    author = "Aaij, Roel and others",
    collaboration = "LHCb",
    title = "{LHCb Detector Performance}",
    journal = "Int. J. Mod. Phys. A",
    volume = "30",
    number = "07",
    pages = "1530022",
    year = "2015"
}

@article{LHCb:2008vvz,
    author = "Alves, Jr., A. Augusto and others",
    collaboration = "LHCb",
    title = "{The LHCb Detector at the LHC}",
    reportNumber = "LHCb-DP-2008-001",
    doi = "10.1088/1748-0221/3/08/S08005",
    journal = "JINST",
    volume = "3",
    pages = "S08005",
    year = "2008"
}

@article{LHCb:2025nob,
    author = "Aaij, Roel and others",
    collaboration = "LHCb",
    title = "{Measurement of the $Z$-Boson Mass}",
    journal = "Phys. Rev. Lett.",
    volume = "135",
    number = "16",
    pages = "161802",
    year = "2025"
}

@article{LHCb:2025msn,
    author = "Aaij, Roel and others",
    collaboration = "LHCb",
    title = "{Measurement of the $W$ {\textrightarrow} {\ensuremath{\mu}}{\ensuremath{\nu}} cross-sections as a function of the muon transverse momentum in $pp$ collisions at 5.02 TeV}",    
    journal = "JHEP",
    volume = "03",
    pages = "148",
    year = "2026"
}

@article{CDF:2022hxs,
    author = "Aaltonen, T. and others",
    collaboration = "CDF",
    title = "{High-precision measurement of the $W$ boson mass with the CDF II detector}",
    journal = "Science",
    volume = "376",
    number = "6589",
    pages = "170--176",
    year = "2022"
}

@article{LHCb:2025kfp,
    author = "Aaij, Roel and others",
    collaboration = "LHCb",
    title = "{Measurement of the top-quark production cross-section and charge asymmetry at LHCb}",
    eprint = "2512.11324",
    journal="submitted to journal",
    archivePrefix = "arXiv",
    primaryClass = "hep-ex",
    reportNumber = "LHCb-PAPER-2025-057, CERN-EP-2025-274",
    month = "12",
    year = "2025"
}

@article{Gauld:2014pxa,
    author = "Gauld, Rhorry",
    title = "{Leptonic top-quark asymmetry predictions at LHCb}",
    eprint = "1409.8631",
    archivePrefix = "arXiv",
    primaryClass = "hep-ph",
    doi = "10.1103/PhysRevD.91.054029",
    journal = "Phys. Rev. D",
    volume = "91",
    pages = "054029",
    year = "2015"
}

@article{Cacciari:2008gp,
    author = "Cacciari, Matteo and Salam, Gavin P. and Soyez, Gregory",
    title = "{The anti-$k_t$ jet clustering algorithm}",
    eprint = "0802.1189",
    archivePrefix = "arXiv",
    primaryClass = "hep-ph",
    reportNumber = "LPTHE-07-03",
    doi = "10.1088/1126-6708/2008/04/063",
    journal = "JHEP",
    volume = "04",
    pages = "063",
    year = "2008"
}

@article{LHCb:2026ezb,
    author = "Aaij, Roel and others",
    collaboration = "LHCb",
    title = "{Machine learning techniques for jet reconstruction at LHCb and application to the search for $H \to b \bar{b}$ and $H \to c \bar{c}$ in $\sqrt{s}=13$ TeV $pp$ collisions}",
    eprint = "2601.16802",
    journal="submitted to journal",
    archivePrefix = "arXiv",
    primaryClass = "hep-ex",
    reportNumber = "LHCb-PAPER-2025-034, CERN-EP-2025-275",
    month = "1",
    year = "2026"
}

@article{LHCb:2026dan,
    author = "Aaij, Roel and others",
    collaboration = "LHCb",
    title = "{Measurement of the $W$-boson production cross-sections in $pp$ collisions at $\sqrt{s}$ = 13 TeV in the forward region}",
    journal="submitted to journal",
    eprint = "2604.12706",
    archivePrefix = "arXiv",
    primaryClass = "hep-ex",
    reportNumber = "LHCb-PAPER-2025-070, CERN-EP-2026-083",
    month = "4",
    year = "2026"
}

@article{LHCb:2026biq,
    author = "Aaij, Roel and others",
    collaboration = "LHCb",
    title = "{Precision measurement of the muon charge asymmetry from $W$-boson decays in $pp$ collisions at $\sqrt{s}$ = 13 TeV in the forward region}",
    journal="submitted to journal",
    eprint = "2604.12593",
    archivePrefix = "arXiv",
    primaryClass = "hep-ex",
    reportNumber = "LHCb-PAPER-2025-071, CERN-EP-2026-084",
    month = "4",
    year = "2026"
}

@article{Barter:2021npd,
    author = "Barter, William and Pili, Martina and Vesterinen, Mika",
    title = "{A simple method to determine charge-dependent curvature biases in track reconstruction in hadron collider experiments}",
    eprint = "2101.05675",
    archivePrefix = "arXiv",
    primaryClass = "hep-ex",
    doi = "10.1140/epjc/s10052-021-09016-9",
    journal = "Eur. Phys. J. C",
    volume = "81",
    number = "3",
    pages = "251",
    year = "2021"
}

@article{LHCb:2023yqm,
    author = "Aaij, R. and others",
    collaboration = "LHCb",
    title = "{Curvature-bias corrections using a pseudomass method}",
    eprint = "2311.04670",
    archivePrefix = "arXiv",
    primaryClass = "hep-ex",
    reportNumber = "LHCb-DP-2023-001, CERN-EP-2023-246",
    doi = "10.1088/1748-0221/19/03/P03010",
    journal = "JINST",
    volume = "19",
    number = "03",
    pages = "P03010",
    year = "2024"
}

@article{ALEPH:2013dgf,
    author = "Schael, S. and others",
    collaboration = "ALEPH, DELPHI, L3, OPAL, LEP Electroweak",
    title = "{Electroweak Measurements in Electron-Positron Collisions at $W$-Boson-Pair Energies at LEP}",
    eprint = "1302.3415",
    archivePrefix = "arXiv",
    primaryClass = "hep-ex",
    reportNumber = "CERN-PH-EP-2013-022",
    doi = "10.1016/j.physrep.2013.07.004",
    journal = "Phys. Rept.",
    volume = "532",
    pages = "119--244",
    year = "2013"
}

@article{LHC-TeVMWWorkingGroup:2023zkn,
    author = "Amoroso, Simone and others",
    collaboration = "LHC-TeV~MW~Working~Group",
    title = "{Compatibility and combination of world $W$-boson mass measurements}",
    eprint = "2308.09417",
    archivePrefix = "arXiv",
    primaryClass = "hep-ex",
    doi = "10.1140/epjc/s10052-024-12532-z",
    journal = "Eur. Phys. J. C",
    volume = "84",
    number = "5",
    pages = "451",
    year = "2024"
}

@article{LHCb:2021bjt,
    author = "Aaij, Roel and others",
    collaboration = "LHCb",
    title = "{Measurement of the $W$ boson mass}",
    eprint = "2109.01113",
    archivePrefix = "arXiv",
    primaryClass = "hep-ex",
    reportNumber = "LHCb-PAPER-2021-024, CERN-EP-2021-170",
    doi = "10.1007/JHEP01(2022)036",
    journal = "JHEP",
    volume = "01",
    pages = "036",
    year = "2022"
}

@article{LHCb:2025gbn,
    author = "Aaij, Roel and others",
    collaboration = "LHCb",
    title = "{Search for resonances decaying to photon pairs with masses between 4.9 and 19.4 GeV}",
    eprint = "2507.14390",
    journal="submitted to journal",
    archivePrefix = "arXiv",
    primaryClass = "hep-ex",
    reportNumber = "CERN-EP-2025-133, LHCb-PAPER-2025-012",
    month = "7",
    year = "2025"
}

@book{Breiman:2017lcz,
    author = "Breiman, Leo and Friedman, Jerome and Olshen, R. A. and Stone, Charles J.",
    title = "{Classification and Regression Trees (1st ed.)}",
    doi = "10.1201/9781315139470",
    publisher = "Chapman and Hall/CRC",
    year = "1984"
}

@article{Freund:1997xna,
    author = "Freund, Yoav and Schapire, Robert E.",
    title = "{A Decision-Theoretic Generalization of On-Line Learning and an Application to Boosting}",
    doi = "10.1006/jcss.1997.1504",
    journal = "J. Comput. Syst. Sci.",
    volume = "55",
    number = "1",
    pages = "119--139",
    year = "1997"
}

@article{LHCb:2015tna,
    author = "Aaij, Roel and others",
    collaboration = "LHCb",
    title = "{Identification of beauty and charm quark jets at LHCb}",
    eprint = "1504.07670",
    archivePrefix = "arXiv",
    primaryClass = "hep-ex",
    reportNumber = "LHCB-PAPER-2015-016, CERN-PH-EP-2015-101",
    doi = "10.1088/1748-0221/10/06/P06013",
    journal = "JINST",
    volume = "10",
    number = "06",
    pages = "P06013",
    year = "2015"
}

@article{LHCb:2025ymr,
    author = "Aaij, Roel and others",
    collaboration = "LHCb",
    title = "{Search for heavy neutral leptons in $B$-meson decays}",
    eprint = "2512.14551",
    archivePrefix = "arXiv",
    primaryClass = "hep-ex",
    reportNumber = "LHCb-PAPER-2025-042, CERN-EP-2025-264",
    doi = "10.1007/JHEP03(2026)178",
    journal = "JHEP",
    volume = "03",
    pages = "178",
    year = "2026"
}

@article{LHCb:2014osd,
    author = "Aaij, Roel and others",
    collaboration = "LHCb",
    title = "{Search for Majorana neutrinos in $B^- \to \pi^+\mu^-\mu^-$ decays}",
    eprint = "1401.5361",
    archivePrefix = "arXiv",
    primaryClass = "hep-ex",
    reportNumber = "CERN-PH-EP-2014-002, LHCB-PAPER-2013-064",
    doi = "10.1103/PhysRevLett.112.131802",
    journal = "Phys. Rev. Lett.",
    volume = "112",
    number = "13",
    pages = "131802",
    year = "2014"
}

@article{LHCB-FIGURE-2025-001,
        author = "Aaij, Roel and others",
      collaboration = "LHCb",
      title         = "{Performance of LHCb as a feebly interacting particles detector}",
      journal       = "LHCb Public figure", 
      year          = "2025",
      url           = "https://cds.cern.ch/record/2923551",
}


\end{document}